\documentclass[preprint,aps,aip]{revtex4-1}
\usepackage{epsfig}
\usepackage{color}
\begin{document}

\title{Micromechanical Model for Self-Organized Impurity Nanorod 
Arrays in Epitaxial YBa$_2$Cu$_3$O$_{7-\delta}$ Films}

\author{Jack J. Shi and Judy Z. Wu}
\affiliation{Department of Physics \& Astronomy, 
The University of Kansas, Lawrence, KS 66045}
\date{June 7, 2011}

\begin{abstract} 
A micromechanical model based on the theory of elasticity has been developed 
to study the configuration of self-assembled impurity nanostructures in high 
temperature superconducting YBa$_2$Cu$_3$O$_{7-\delta}$ films. With the 
calculated equilibrium strain and elastic energy of the impurity doped film, 
a phase diagram of lattice mismatches $vs.$ elastic constants of the dopant 
was obtained for the energetically-preferred orientation of impurity nanorods. 
The calculation of the nanorod orientation and the film lattice deformation 
has yielded an excellent agreement with experimental measurements.

\end{abstract}

\maketitle

\noindent
{\bf I. Introduction}

Self-organization of nanostructures in epitaxial films may provide a 
unique approach to design and tailor physical properties of 
nano-composite films by controlling the morphology of the 
nanostructures. Impurity doped high temperature superconducting (HTS)
YBa$_2$Cu$_3$O$_{7-\delta}$ (YBCO) film is an excellent example among 
other oxide nano-composites. 
In pursuing electronic and electric applications of HTS materials, 
extensive efforts have been taken to improve the critical current 
density ($J_c$) by incorporating impurity nanostructures as strong 
magnetic vortex pinning centers in YBCO films \cite{Goyal1,Foltyn}. 
Self-organized impurity nanostructures, such as BaZrO$_3$ (BZO) nanorod 
(NR) array oriented along the YBCO $c$ axis, were found to dramatically 
improve $J_c$ with up to a fivefold increase in applied magnetic field 
up to 5 T \cite{Kang}. The strong correlated pinning provided by the 
aligned NRs also reduces the $J_c$ anisotropy in the applied field 
direction. BZO has, however, a relatively large lattice mismatch with 
YBCO, resulting in a reduced critical temperature due to a seriously 
strained YBCO lattice upon BZO doping \cite{Claudia}. While a strained 
lattice is essential to the formation of the nanostructures, a 
quantitative control of the strain at a microscopic scale could optimize 
physical properties of nano-composite films. It is thus desirable to 
have quantitative criteria in selecting dopants for a given film matrix 
based on an understanding of the strain field and its role in the 
formation of the nanostructures. Such criteria are, however, not 
currently available. Moreover, the perfectly aligned NRs along the $c$ 
axis in YBCO films is not ideal since the in-field $J_c$ is not much 
improved or even reduced when the applied field is not along the $c$ 
axis. A three-dimensional pinning landscape in the film could be a 
solution to this problem and some attempts have succeeded recently in 
generating splayed BZO NRs using either vicinal substrate \cite{Baca2} 
or secondary impurity nanoparticles \cite{Maiorov}. In both cases, an 
additional lattice mismatch is introduced into the film matrix to change 
the strain field in the film. Theoretical understanding of the interplay 
of strains due to multiple mismatched lattices becomes essential to the 
control of the microstructure and physical properties of nano-composite 
films. Moreover, a better modeling of the strain field in the doped YBCO 
films could also help to understand the change of the superconducting 
properties of the films due to the doping \cite{Rodriguez}.

Many theoretical studies have been made on self-organization of 
nanostructures in crystals \cite{Khachaturyan,Bimberg}. The analysis 
based on the elastic theory of strained lattices has been successful 
on the self-organization of elastic stress domains and nanostructures 
\cite{Khachaturyan,Bimberg,Roitburd,Bruinsma,Ipatova,Ledentsov}. The 
elastic strain models developed in those studies consider the strain due 
to a mismatch between two different lattices with a coherent interface. 
Recently, extensive studies have also been made on multiferroic 
nanostructures in epitaxial composite films using phase field model 
\cite{Artemev,Slutsker1,Slutsker2}, in which a few phenomenological model 
parameters need to be properly selected for modeling the strain field 
in multiple mismatched lattices. This paper presents a micromechanical 
model based on the theory of elasticity with experimentally measured 
elastic constants and without any fitting paramenter for the strain 
field in epitaxial nano-composite films with impurity nanostructures. 
In this model, the boundary conditions of the equilibrium equation at 
interfaces are more properly specified as compared with the previous 
models and, therefore, no phenomenological fitting parameter is 
needed. This model therefore represents a considerable improvement 
from the previous models. While the doped YBCO epitaxial film is 
studied as an example system, the method of the strain analysis 
developed here can be readily generalized to study the 
self-organization of nanostructures in other epitaxial nano-composite 
films. This paper is organized as follows. In Sec. II, the basic 
formalism of the model is discussed. Studies of the orientation of 
impurity NRs in epitaxial films on lattice-matched and -mismatched 
substrates are presented in Sec. III and IV, respectively. Section V 
contains a conclusion. 

\vspace{0.1in}
\noindent
{\bf II. Elasticity Model of Impurity Nanostructures in Epitaxial Films}

In this model, the formation of aligned impurity NRs in epitaxial 
films is assumed to be the consequence of the relaxation to the 
energetically-preferred elastic equilibrium of coherently strained 
lattices due to lattice mismatches among film, dopant, and substrate. 
The assumption of the coherently strained lattices is based on the 
following considerations. (a) With a small volume density of dopant, the 
film and dopant lattices are likely to be coherently strained as long 
as the lattice mismatch between film and dopant is sufficiently small. 
Experimentally, semi-coherent interface between YBCO and dopant with 
some dislocations has been observed \cite{Claudia}. The effect of the 
dislocations is neglected here in order to obtain an analytical solution. 
(b) In the case of the film on a lattice-mismatched substrate, as long 
as the lattice mismatch is sufficiently small, the film could be 
coherently strained near the film-substrate interface until a critical 
thickness is reached. Beyond this coherent region, the strain in the 
film can be released by introducing dislocations. Since the crystalline 
configuration of impurity nanostructures in epitaxial films is strongly 
influenced by their initial formation near the film/substrate interface, 
only the strain field in the coherent region of the film is considered 
in this study and the effect of non-coherent over-layers is neglected. 
Therefore, the possibility to fabricate impurity NRs with the vertical 
or horizontal alignment in $c$-oriented films depends on the elastic 
energy of the strained lattices with respect to other possible NR 
configurations. Note that the horizontal and vertical alignments are the 
most probable, if not only, configurations assuming dopant-film epitaxy 
is maintained. On the (001)-cut SrTiO$_3$ (STO) substrate, the YBCO 
$ab$ planes are considered to be twinned. This allows simplification 
of the three-dimensional system to a two-dimensional one that contains 
the [100] and [001] directions. Note that all the calculations can be 
readily extended to the three-dimensional case of orthorhombic lattices. 
Let $x_1$ and $x_3$ be the components of the two-dimensional coordinate 
along the [100] and [001] direction, respectively. The elastic energy 
of strained lattices can be determined by 
\begin{equation}
\label{E_el}
E_{el}=\int\limits_{film}E_1\,dV+\int\limits_{dopant}E_2\,dV
      +\int\limits_{substrate}E_3\,dV
\end{equation}
where $E_i$ is the elastic energy density for film ($i=1$), dopant 
($i=2$), and substrate ($i=3$), respectively. Considering the 
tetragonal symmetry of twinned YBCO film and the cubic symmetry of 
substrate and dopant, the elastic energy density can be written as 
\cite{Landau}
\begin{equation}
\label{E_YBCO1}
E_i=\frac{1}{2}\lambda_{i1} u_{11}^2+\frac{1}{2}\lambda_{i2} u_{33}^2  
    +\lambda_{i3} u_{11}u_{33}+\lambda_{i4} u_{13}^2 
\end{equation} 
where $u_{jk}$ is the strain tensor and $\lambda_{i1}=c^{(i)}_{11}$, 
$\lambda_{i2}=c^{(i)}_{33}$, $\lambda_{i3}=c^{(i)}_{13}$, and 
$\lambda_{i4}=c^{(i)}_{55}$ are the elastic constants of the material 
labeled with $i$. The elastic constants used in this study can be 
found in Refs. \cite{Kuzel,Dieguez,Bouhemadou,Poindexter}. Note that 
the interaction energies at film/substrate and film/dopant interfaces 
are included implicitly in $E_{i}$ as $u_{jk}$ is the solution of 
equilibrium equations with the boundary conditions that are the result 
of the interface interactions. Using the general summation rule (also 
in other formulas in this paper), the equilibrium equations can be 
written as \cite{Landau}
\begin{equation}
\label{equilibrium}
\frac{\partial }{\partial x_k}
\left(\frac{\partial E_i}{\partial u_{jk}} \right)=0  \,,
\end{equation}
where $\sigma_{jk}=\partial E_i/\partial u_{jk}$ is the stress tensor. 
At an interface between two coherently bonded lattices, the boundary 
condition of Eq. (\ref{equilibrium}) prescribes continuity of the force 
on the interface and allows for a discontinuity of the strain across 
the interface. Let $\vec{n}=(n_1,\,n_3)$ and $\vec{s}=(s_1,\,s_3)$ be 
the unit vectors normal and tangential to an interface. The boundary 
condition for the continuity of the force on an interface is
\begin{equation}
\label{boundary1}
 n_k \left[\sigma_{jk}(1)-\sigma_{jk}(2)\right]=0 \,, 
\end{equation}
where $\sigma_{jk}(1)$ and $\sigma_{jk}(2)$ are the stress at the 
interface in lattice $1$ and $2$, respectively. Along the tangential 
direction of an interface, the lattices on the two sides are deformed 
in order to match each other under the interaction between lattices. 
As illustrated in Fig. 1a, if the lattice at one side of the interface 
is stretched (compressed) the lattice at the other side is compressed 
(stretched) accordingly. The total deformation in length of the two 
lattices along the tangential direction is thus assumed to equal the 
difference of their mismatched lattice constants \cite{Roitburd}. Note 
that the underlying assumption here is that the bonding between two 
lattices (such as adsorbate-substrate interaction) is much stronger than 
the lattice bounding in each material. This assumption is reasonable for 
the case of the impurity doped YBCO films and many other epitaxial films 
when lattice mismatch is small or the defects caused by lattice mismatch 
is not a major concern. As the normal components ($u_{kk}$) of a strain 
tensor describe the change in length per unit length, the discontinuity 
of the normal strain component in the tangential direction of an 
interface is 
\begin{equation}
\label{boundary2}
s_k\left[u_{kk}(1)-(1+f_k)\,u_{kk}(2)-f_k\right]=0 \,,
\end{equation}
where $f_1=a_2/a_1-1$ and $f_3=c_2/c_1-1$ are the lattice mismatches 
at the interface along the [100] and [001] direction, respectively, 
and $(a_1,c_1)$ and $(a_2,c_2)$ are the lattice constants at each side 
of the interface. It should be noted that if the above assumption of 
a strong bonding between two mismatched lattices is invalid, the 
left-hand side of Eq. (\ref{boundary2}) will no longer equal to zero.
In that case, Eq. (\ref{boundary2}) could be modified by replacing zero 
with a parameter that has to be determined empirically or by model 
fitting. Moreover, as illustrated in Fig. 1b, the elastic forces from 
the two deformed lattices that exert on a surface element perpendicular 
to the tangential direction of the interface should be balanced at the 
interface, $i.e.$ 
\begin{equation}
\label{boundary3}
s_k\left[\sigma_{jk}(1)+\sigma_{jk}(2)\right]=0 \,.
\end{equation}
Finally, the strain vanishes deep inside the substrate and the boundary 
conditions at the top surface of a film is simply $n_k\sigma_{jk}=0$. 
It should be noted that in this model the lattice mismatches between 
materials are treated locally at each interface, which allows the 
consideration of different lattice mismatch at different interface. 
This set of boundary conditions is more properly specified than that 
used in the previous elastic models \cite{Bimberg,Artemev,Slutsker2}. 
In general, the solution of PDE in Eq. (\ref{equilibrium}) is not 
unique unless its boundary condition is properly specified. With 
the boundary conditions in Eqs. (\ref{boundary1})-(\ref{boundary3}), 
Eq. (\ref{equilibrium}) for epitaxial films with impurity NRs 
becomes a well-posed PDE problem as the solution is unique and, 
therefore, no phenomenological fitting parameter is needed in this 
model. As far as we know this set of the boundary conditions has not 
been explicitly used before for studying the strain field in composite 
materials. By solving equilibrium equation in each material with the 
boundary conditions, the equilibrium strain for a given configuration 
of impurity nanostructures in a film matrix can be obtained and the 
elastic energy can then be calculated using Eqs. (\ref{E_el}) and 
({\ref{E_YBCO1}). In order to test this model, an experimentally 
well-studied case was examined \cite{XXX}. In the experiment, a few 
monolayers of the $c$-oriented SrLaAlO$_4$ were deposited on STO 
substrate, where the lattice mismatch between the film and substrate 
is about 4\%. The model calculated SrLaAlO$_4$ lattice deformation 
of 1.0\% expansion in the $ab$ plane and 1.0\% contraction along the 
$c$ axis agrees well with the experimental measurement \cite{XXX} of 
about $1.5\%$ expansion and contraction in the respective directions.

\vspace{0.1in}
\noindent
{\bf III. Impurity Nanorods in Epitaxial Films on Lattice-Matched 
Substrates}

For the case of impurity doping in YBCO films, let's first consider  
aligned impurity NRs in the $c$-oriented YBCO film on (001)-cut STO 
substrate. To determine the energetically-preferred orientation of NRs, 
the elastic energies for the configurations of NRs aligned in the [001] 
or [100] direction in the film were calculated and compared. Since the 
lattice mismatch between the twinned YBCO $ab$ planes and STO is 
negligible, only the strain field due to the lattice mismatch between 
film and dopant needs to be considered. In order to solve $u_{jk}$ 
analytically, the system was further simplified as follows. (a) The 
length of the NR is much larger than its diameter so that the effect 
of the interfaces between film and dopant at the two ends of a NR is
negligible. Hence, the strain field is approximately uniform in the
direction along NRs. (b) The volume density of NRs is small so that 
the strain decays to zero at the halfway point between two neighboring 
NRs. (c) NRs are assumed to be uniformly distributed in the film.
Figure 2 illustrates this geometric configuration of NRs in epitaxial 
films for the solution of Eq. (\ref{equilibrium}). With these 
simplifications, $u_{ij}$ was found to be linearly dependent on the 
coordinates for both the configurations of the NR orientation. When 
NRs align in the [001] direction, for example, the equilibrium strain 
in YBCO was obtained as 
\begin{equation}
\label{solution1}
\left\{\begin{array}{l}
 u_{11} =-A\lambda_{13}(1-x_1/D) \\  
 u_{33} =-\left(\lambda_{11}/\lambda_{13}\right)u_{11} \\
 u_{13} = 0 
\end{array}\right. 
\end{equation} 
where $x_1\in[0,D]$ with $x_1=0$ at the center of a NR and $x_1=D$ at 
the halfway point between two neighboring NRs (see Fig. 2a), 
\begin{equation}
\label{A}
A=\frac{w_2\, f_3}{(1-\rho)\left[\lambda_{11}w_2
                     +(1+f_3)\lambda_{12}w_1\right]}\,,
\end{equation}
\begin{equation} 
\label{wi}
w_i=\frac{\lambda_{i1}\lambda_{i2}-\lambda_{i3}^2}{\lambda_{i2}} \,,
\end{equation}
$\rho$ is the volume density of NRs, and $f_3$ and $f_1$ are the 
lattice mismatches between the film and dopant in the [001] and [100] 
direction, respectively. For the twined YBCO $ab$-planes, the $a$-axis 
lattice constant in $f_1$ is the average of the lattice constants of 
the $a$ and $b$ axis. A similar solution was also obtained for the 
equilibrium strain in the configuration of NRs aligned in the [100] 
direction. With the obtained equilibrium strains, the elastic energy 
difference between the two orientation configurations was calculated as
\begin{widetext}
\begin{equation}
\label{dE1}
E_{el}[100]-E_{el}[001]
 =\frac{V\,\lambda^2_{11}\,w_1\,w_2^2\,f_3^2}
 {6\left[\lambda_{11}w_2+\lambda_{12} w_1(1+f_3)\right]^2}
 \left[\left(1-\rho+\frac{w_1}{w_2}\rho\right)G
-\frac{w_1\lambda_{12}(\lambda_{11}-\lambda_{12})}
      {\lambda^2_{11}w_2}\rho \,\right] 
\end{equation}
\end{widetext}
where $E_{el}[100]$ and $E_{el}[001]$ are the elastic energies of the 
film with NRs aligned in the $[100]$ and $[001]$ directions, respectively, 
$V$ is the volume of the film, and for $f_1<<1$ and $f_3<<1$,
\begin{equation}
\label{phase}
G=\left[\frac{\lambda_{12}w_1+\lambda_{11}w_2}
             {\lambda_{11}(w_1+w_2)}\right]^2 
        \left(\frac{f_1}{f_3}\right)^2
       -\frac{\lambda_{12}}{\lambda_{11}}\,.
\end{equation}  
Considering $\rho<<1$, $G$ as a function of $w_2$ and $|f_1/f_3|$ can 
be conveniently used as a state function for the NR orientation, where
$w_2$ is of the elastic constants of dopant. Note that 
$0<w_i<\lambda_{i1}$ for a positive-definite elastic energy. When 
$G\left(w_2,\,\left|f_1/f_3\right|\right)<0$, $E_{el}[100]-E_{el}[001]<0$ 
and it is not possible to have NRs aligned in the [001] direction and 
vice versa. Hence, $G\left(w_2,\,\left|f_1/f_3\right|\right)=0$ yields 
a phase boundary that separates the regions in the parameter space of 
$\left(w_2,\,\left|f_1/f_3\right|\right)$ for dopant, where the 
vertical or horizontal alignment of NRs is not possible in the 
$c$-oriented film on a lattice-matched substrate. Figure 3 plots this 
phase diagram for the doped YBCO film where the vertical alignment 
of NRs is only possible in region I. For the BZO or BSO NRs, as shown 
in the figure, the vertical alignment is the energetically preferred 
state in the $c$-oriented YBCO film on a lattice-matched substrate, 
which is consistent with experimental observations 
\cite{Baca2,Mele,Varanasi}. The preference of the NR orientation is 
determined by the difference in the lattice mismatches between the 
film and dopant in the [001] and [100] directions and the anisotropy 
of the elastic constants. Because $f_3<f_1$, the alignment of NRs in 
the [001] direction results in less deformation in the YBCO lattice. 
Since $\lambda_{12}<\lambda_{11}$, moreover, the film lattice along the 
[001] direction is relatively easier to deform and, therefore, easier 
to accommodate NRs. The strain energy is therefore lower when NRs 
aligns in the [001] direction. Recently, efforts have also been made 
to fabricate the YBCO films with the vertically-aligned Y$_2$O$_3$ or 
CeO$_3$ NRs but have not been successful. As shown in Fig. 3, Y$_2$O$_3$ 
and CeO$_3$ are all in region II of the phase diagram, which confirms 
the negative result of the experiments. It should be emphasized that 
the condition obtained here is only applicable to the case of the 
impurity NRs in the film without significant coexistence of other forms 
of impurity inclusions. The existence of secondary impurity inclusions 
could substantially change the strain field in the film and could in 
turn alter the elastic energy minima. This may explain the observed BZO 
NRs splay with addition of Y$_2$O$_3$ nanoparticles in YBCO films 
\cite{Mele,Maiorov}. 

\vspace{0.1in}
\noindent
{\bf IV. Impurity Nanorods in Epitaxial Films on Lattice-Mismatched 
Substrates}

In the case of impurity-doped YBCO films on lattice-mismatched substrates, 
the lattice mismatch between the film and substrate results in a strained 
film lattice, which in turn changes the lattice mismatch between the 
film and dopant. Considering that the film/substrate mismatch is the 
same in both the directions of the interface or only in one direction,
the strain in the film due to the film/substrate lattice mismatch can 
be easily solved from Eq. (\ref{equilibrium}) as
\begin{equation}
\label{substrain}
\left\{\begin{array}{l}
u_{11}=\Gamma (1-x_3/h)  \\
u_{33}=-\xi \left(\lambda_{13}/\lambda_{12}\right)u_{11} \\
u_{12}=0
\end{array}\right.
\end{equation}
where $x_3\in[0,\,h]$ with $x_3=0$ at the substrate surface, $h$ is the 
film thickness, 
\begin{equation}
\label{Bconst}
\Gamma=\frac{w_3f_s}{w_3+w_1(1+f_s)}\,,
\end{equation} 
$w_i$ is given in Eq. (\ref{wi}), $f_s=a_3/a_1-1$, $a_3$ is the 
unstrained lattice constant of the substrate, and $\xi=1$ or 2 for the 
cases of the lattice mismatch in one or two directions of the interface, 
respectively. If $f_s>0$, $i.e.$ the substrate lattice is bigger than 
the film lattice, $u_{11}>0$ and $u_{33}<0$, and vise versa. The tensile 
(compressive) strain in the film $ab$ planes due to a mismatched 
substrate leads to a compressive (tensile) strain along the film $c$ axis. 

For the inclusion of impurity NRs in a film on a mismatched substrate, 
the effect of the substrate can be studied approximately by considering 
the NRs in a pre-strained film matrix due to the mismatched substrate. 
Near the film-substrate interface, the changes of the film lattice 
constants due to the mismatched substrate are 
$\delta a_1/a_1\simeq u_{11}(0)=\Gamma$ and 
$\delta c_1/c_1\simeq u_{33}(0)=-\xi(\lambda_{13}/\lambda_{12})\Gamma$. 
Consequently, the lattice mismatch between the film and dopant becomes 
$f_1+\delta f_1$ and $f_3+\delta f_3$, where 
\begin{equation}
\label{delta_f}
\left\{\begin{array}{l}
\delta f_1=-(1+f_1)\Gamma \\
\delta f_3=\xi(1+f_3)(\lambda_{13}/\lambda_{12})\Gamma
\end{array}\right.
\end{equation}
and the state function $G(w_2,\,|f_1/f_3|)$ in Eq. (\ref{phase}) is 
modified as $G(w_2,\,|f_1+\delta f_1|/|f_3+\delta f_3|)$. Figure 2 plots 
$G(w_2,\,|f_1+\delta f_1|/|f_3+\delta f_3|)$ as a function of the 
film/substrate lattice mismatch for cases of BZO and BSO in YBCO on an 
example cubic-lattice substrate that has similar elastic constants to STO. 
It shows that the nanorods align with the $c$ axis (or in the $ab$ plane) 
of the film if $f_s$ is smaller (or larger) than a threshold. When $f_s$ 
is larger enough, the lattice mismatch between the film and dopant is 
substantially altered by the strain in the film due to the mismatched 
substrate and, consequently, the energetically-preferred NR alignment 
changes from the [001] to [100] direction. For both cases of BZO and 
BSO, as shown in Fig. 2, this threshold for the transition of the NR 
orientation is $f_{sc}\simeq+1.2\%$ or $+2.5\%$ for the mismatch in the 
both or only one lattice direction of the substrate surface.

The deformation of the film lattice due to the inclusion of the NRs can 
be calculated by averaging the principal components of the equilibrium 
strain over the film, which can be compared with experimental measurement. 
Because of the different NR orientation, the lattice deformation is 
different in the regions of $f_s<f_{sc}$ and $f_s>f_{sc}$. When 
$f_s<f_{sc}$, the deformation of the film lattice calculated with 
$\rho<<1$ is 
\begin{equation}
\label{Da3}
\begin{array}{l}
{\displaystyle
\frac{\delta a_1}{a_1}=-\frac{\lambda_{13}w_2(f_3+\delta f_3)}
      {2\,[\lambda_{11}w_2+(1+f_3+\delta f_3)\lambda_{12}w_1]} 
      +\frac{1}{2}\Gamma }     
   \vspace{0.1in} \\
{\displaystyle
\frac{\delta c_1}{c_1}=\frac{\lambda_{11}w_2(f_3+\delta f_3)}
      {2\,[\lambda_{11}w_2+(1+f_3+\delta f_3)\lambda_{12}w_1]}
      -\frac{\lambda_{13}}{2\lambda_{12}}\Gamma  }
\end{array}
\end{equation}      
For $f_3>0$, $\delta a_1/a_1<0$ and $\delta c_1/c_1>0$, which 
represents a compression and expansion of the film lattice along the 
$a$ and $c$ axis, respectively. From Eq. (\ref{Da3}), when $f_s=0$, 
$(\delta a_1/a_1)/(\delta c_1/c_1)=-\lambda_{13}/\lambda_{11}$, $i.e.$ 
the ratio of the lattice-constant changes depends only on the elastic 
constants of the film and is independent of the properties of dopant. 
In the case of BZO NRs in YBCO films on the (001)-cut STO substrate, 
it was measured that the YBCO $c$ axis expands about $\sim 0.1\%$ 
\cite{Baca2} while the calculation with Eq. (\ref{Da3}) yields an 
expansion of $0.2\%$. The small discrepancy here may be attributed to 
the presence of dislocations around the BZO NRs, which could release the 
local strain partially \cite{Goyal}. When $f_s>f_{sc}$, the deformation 
of the film lattice becomes
\begin{equation}
\label{Da4}
\frac{\delta a_1}{a_1} 
 =\frac{1}{2}\left[ \frac{w_2(f_1+\delta f_1)}{w_2+(1+f_1+\delta f_1)w_1}
  +\Gamma \right]
\end{equation}
and $\delta c_1/c_1=-(\lambda_{13}/\lambda_{12})(\delta a_1/a_1)$. In 
this case, the film lattice is expanded in the $ab$ planes and 
compressed along the $c$ axis. The ratio of the lattice deformations 
$(\delta a_1/a_1)/(\delta c_1/c_1)$ also depends only on the elastic 
constants of the film.  

\vspace{0.1in}
\noindent
{\bf V. Summary and Discussion}

A micromechanical model based on the theory of elasticity has been 
developed to study the configurations of the self-organized impurity 
nanostructures in epitaxial films. By treating lattice mismatch locally
at interfaces, the strain field due to multiple mismatched lattices 
of film, dopants, and substrate can be simultaneously considered. 
Including the effect of multiple lattice mismatches is important to 
the understanding of impurity nanostructures in nano-composite films 
on lattice-mismatched substrates. With a simplified geometry of 
impurity nanorods in YBCO epitaxial films, the strain field in the 
film was solved in closed form. Based on the analytically calculated 
elastic energy, quantitative criteria have been developed in selecting 
dopants in the YBCO film for possible orientation of the nanorods. 
An excellent agreement between the theoretical predictions and the 
experimentally observed nanostructures has been achieved. The 
significance of the agreement between the model and experiment is that 
there is no any parameter in the model that can be adjusted to fit the 
calculation to the experiment. The success of this model suggests that 
the strain field in coherent interface layers of lattice-mismatched 
dopant, film and substrate is the dominant driving force for the 
self-organization of the nanostructures and fine tuning of the strain 
field through engineering selected interface can lead to controllable 
growth of the desired nanostructures in nano-composite films. This 
micromechanical model can be used as the foundation for a numerical 
study of the impurity nanostructures in presence of dislocations and 
secondary impurity doping in nano-composite films. 

One interesting result of this study is a simple scaling factor 
between the strains in the directions parallel with and perpendicular 
to an interface of two mismatched lattices. As it can be seen in 
Eqs. (\ref{solution1}) and (\ref{substrain}), $u_{33}/u_{11}$ in a 
strained film lattice depends only on the elastic constants of the film 
and is independent of the lattice mismatch or the elastic properties of 
the other mismatched lattice. In the case of strained epitaxial films 
on lattice-mismatched substrates, this phenomenon has been observed 
in an experiment \cite{XXX}. Experimental confirmation of such a 
scaling behavior of the lattice deformation in the nano-composite films 
is important and will provide further insights in the strain mediated 
self-origanization of nanostructures in epitaxial films. Such a study 
could lead to a better understanding and a more accurate modeling of 
the stain field in the nano-composite films.

\vspace{0.1in}
\noindent
{\bf ACKNOWLEDGMENTS}

This work is supported by NSF and ARO under contract no. NSF-DMR-0803149,
NSF-DMR-1105986, NSF EPSCoR-0903806, and ARO-W911NF-0910295.

\newpage

\begin{center} 
{\bf Figure Captions}
\end{center}

\begin{center}
\includegraphics[width=65mm,angle=0]{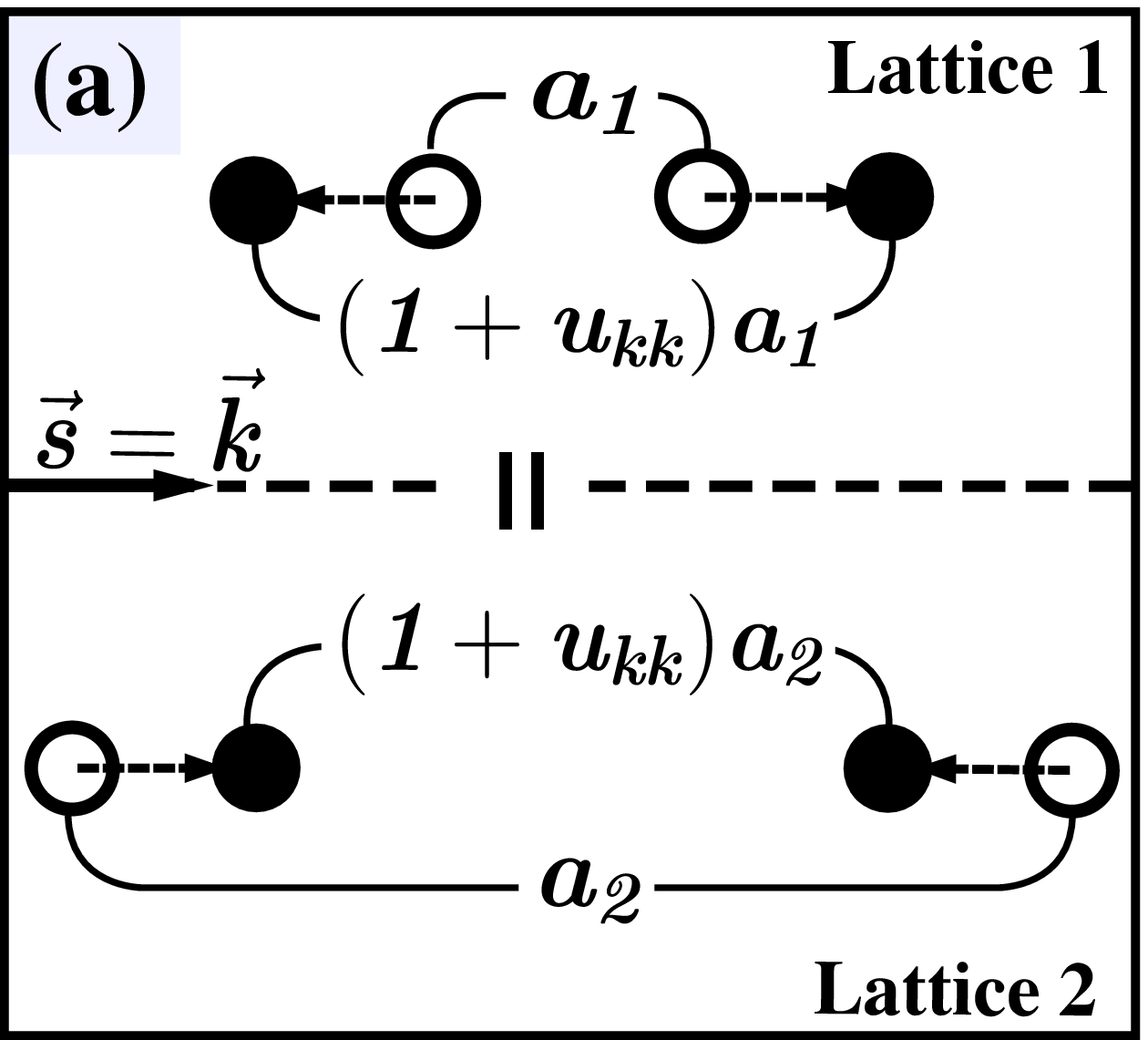}
\hspace{0.2in}
\includegraphics[width=65mm,angle=0]{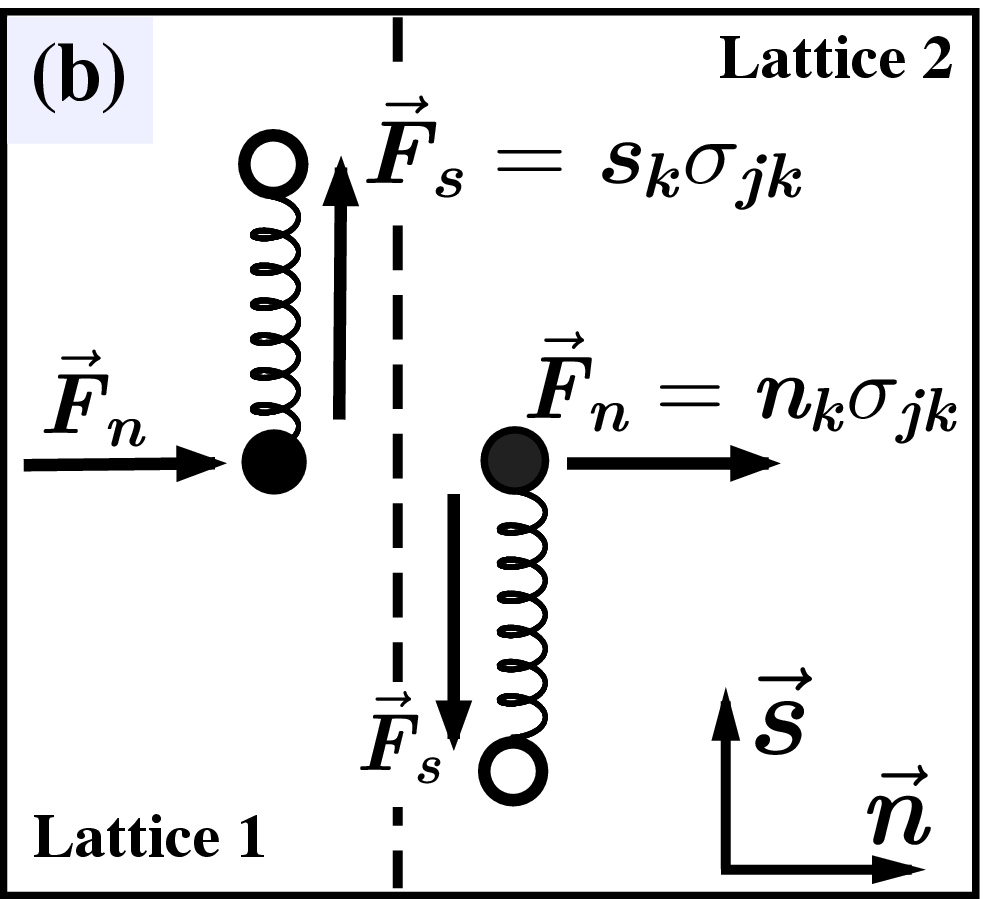}

\vspace{0.2in}
\begin{minipage}{6.0in}
\noindent
Figure 1. 
Illustration of the boundary conditions of Eq. (\ref{equilibrium}) in
(a) Eq. (\ref{boundary2}) and (b) Eqs. (\ref{boundary1}) and 
(\ref{boundary3}). The open and solid circles are the undeformed and 
deformed lattices, respectively, $a_1$ and $a_2$ are the natural 
lattice constants of the two lattices, $u_{kk}$ are the principal 
strain along the interface (dashed line), and $\vec{F}_n$ and 
$\vec{F}_s$ are the elastic forces along the normal ($\vec{n}\,$) and 
tangential ($\vec{s}\,$) direction of the interface, respectively.
  
\end{minipage}
\end{center}

\vspace{0.3in}
\begin{center}
\includegraphics[width=65mm,angle=0]{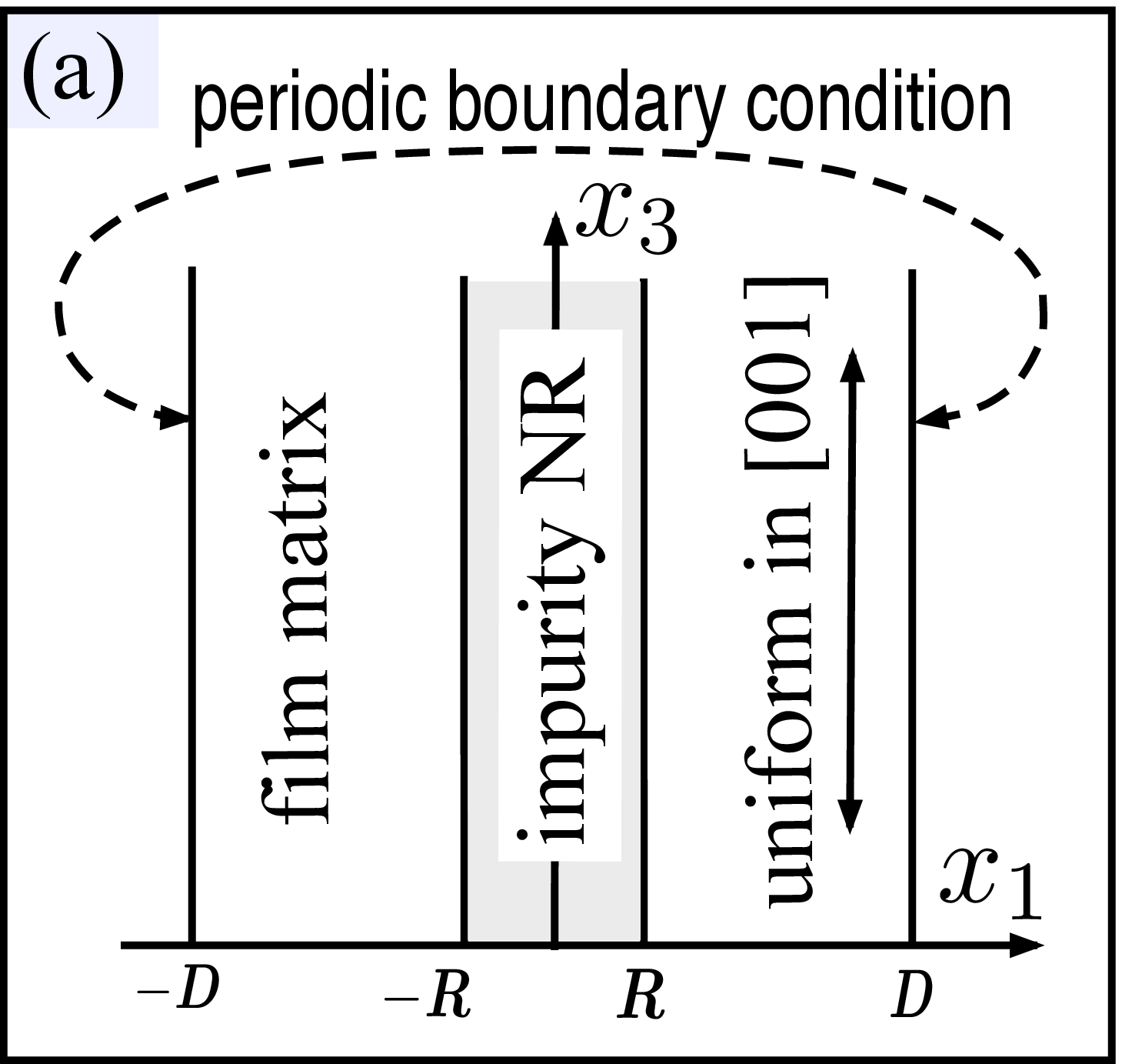}
\hspace{0.2in}
\includegraphics[width=65mm,angle=0]{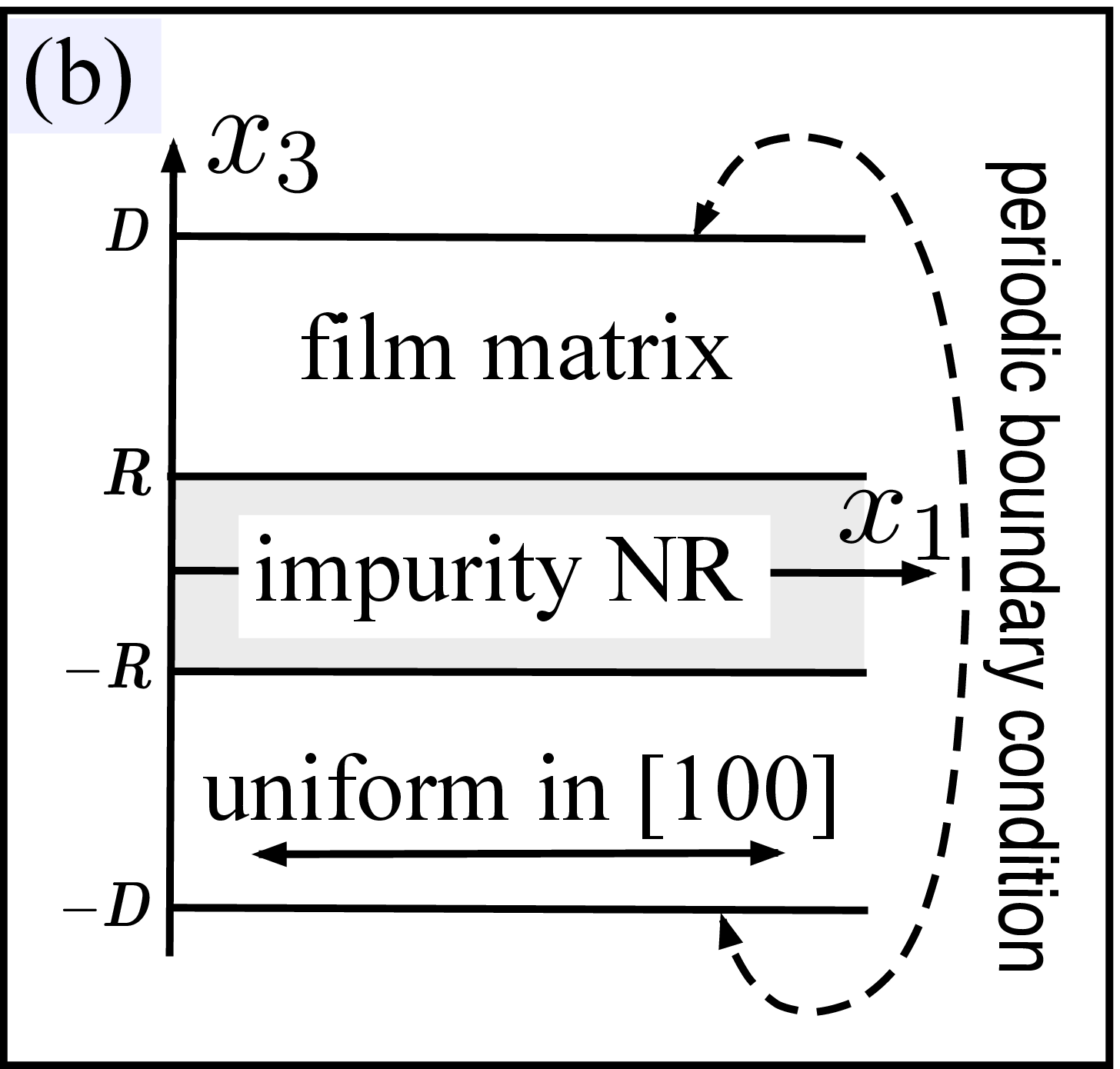}

\vspace{0.2in}
\begin{minipage}{6.0in}
\noindent
Figure 2. 
Geometric configurations of NRs aligned in (a) the $[001]$ direction 
($x_3$-axis) and (b) the $[100]$ direction ($x_1$-axis) for solving  
Eq. (\ref{equilibrium}). $R$ is the radius of the NR and $2D$ is the 
distance between two neighboring NRs.

\end{minipage}
\end{center}

\newpage
\begin{center}
\includegraphics[width=80mm,angle=-90]{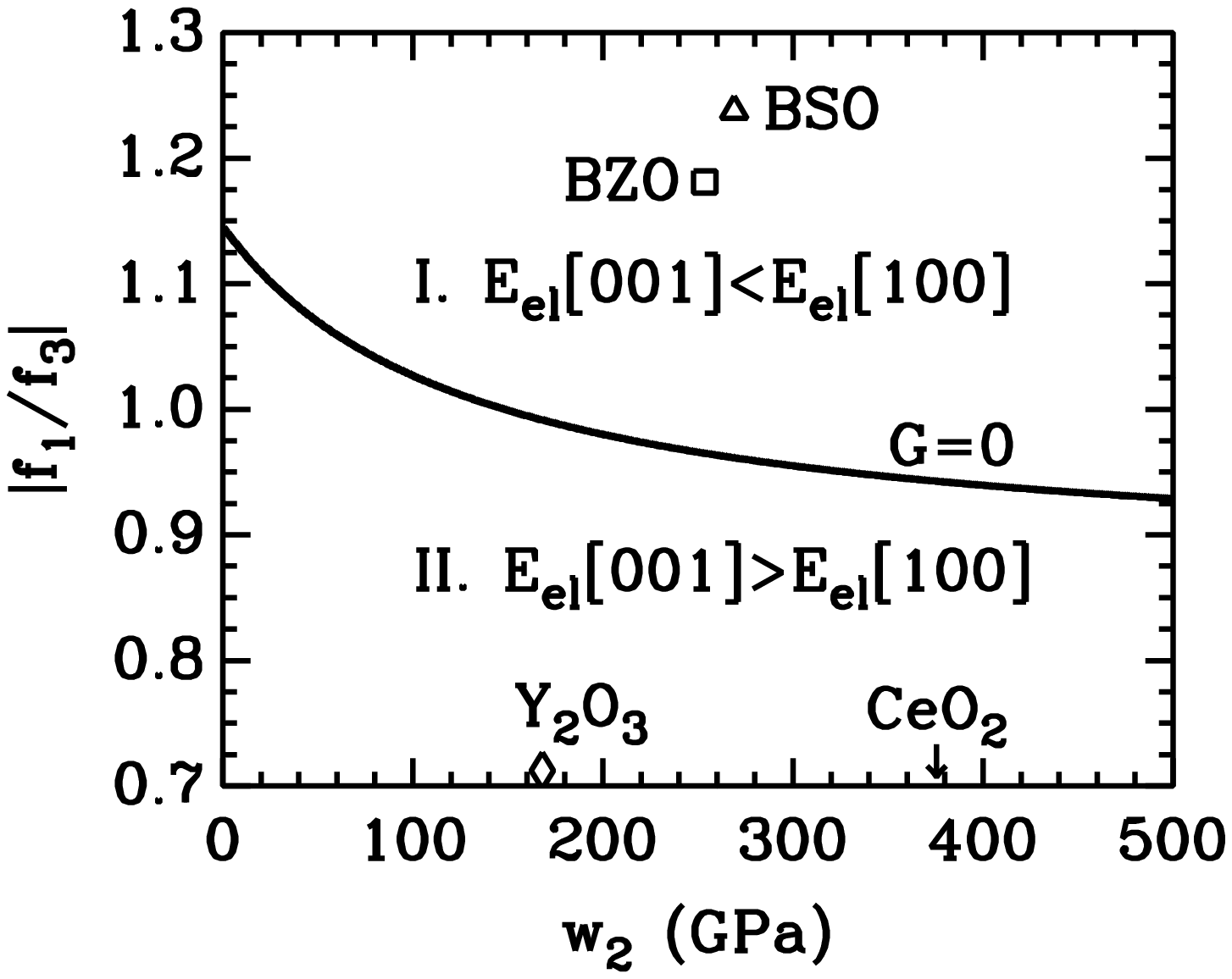}

\vspace{0.2in}
\begin{minipage}{6.0in}
\noindent
Figure 3. 
Threshold of $\left|f_1/f_3\right|$ as a function of $w_2$ for the 
NR orientation in the $c$-oriented YBCO film on lattice-matched 
substrates. Below (above) the solid curve, the vertical (horizontal) 
alignment is unattainable. The square, triangle, and diamond are of 
BZO, BSO, and Y$_2$O$_3$, respectively. The point for CeO$_2$ is 
below the bottom of the figure.
\end{minipage}
\end{center}

\newpage
\vspace{0.2in}
\begin{center} 
\includegraphics[width=80mm,angle=-90]{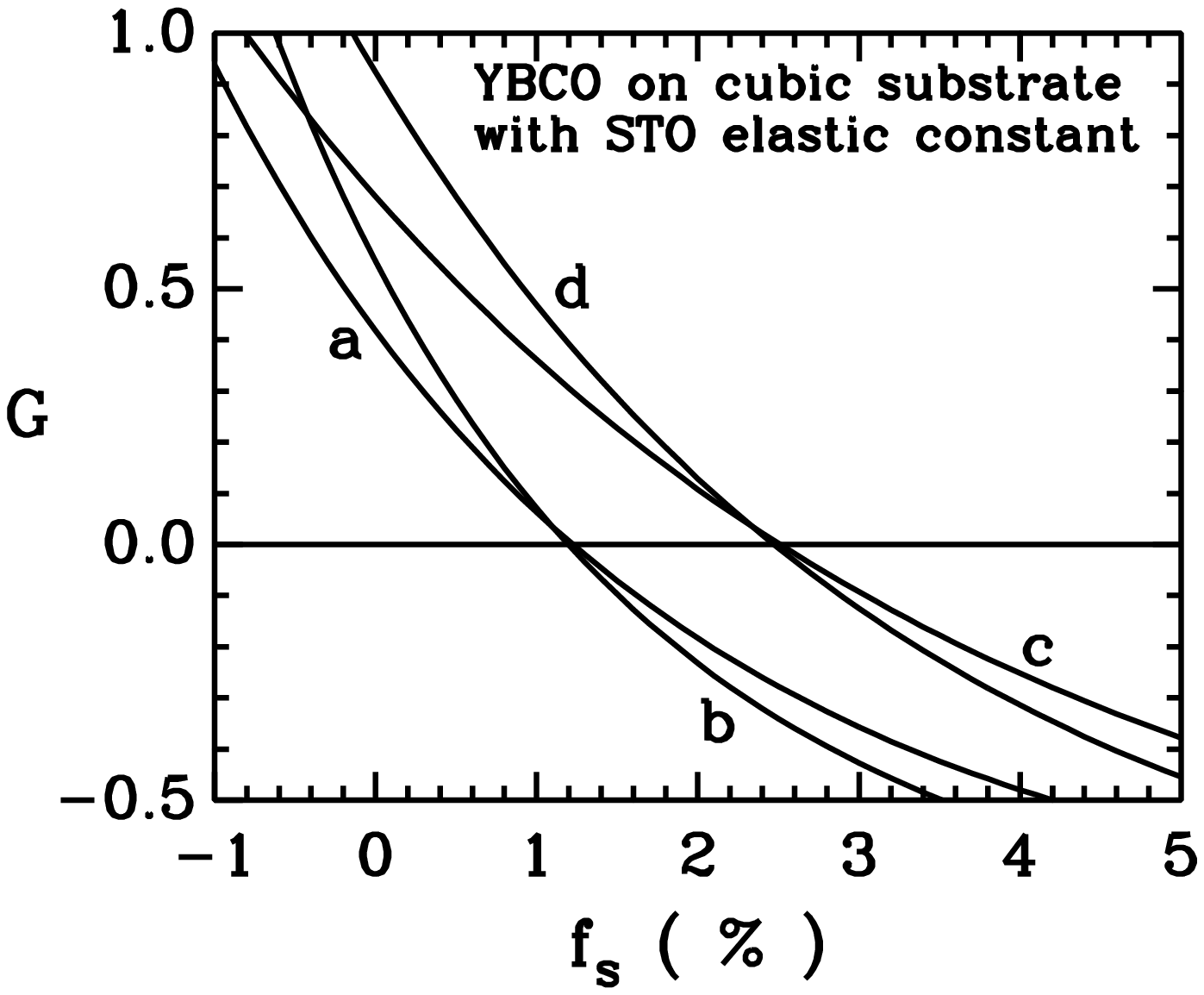}

\vspace{0.2in}
\begin{minipage}{6.0in}
\noindent
Figure 4. 
$G$ $v.s.$ the film/substrate lattice mismatch for BZO (a and c) and BSO 
(b and d) NRs in YBCO film on a cubic-lattice substrate with STO elastic 
constants. (a,b) The mismatch is the same in the [100] and [010] 
directions and the film $ab$ planes are twinned. (c,d) The mismatch is 
only in the [100] direction and the film is detwinned.
\end{minipage}
\end{center}


\begin{thebibliography}{99}
\bibitem{Goyal1}
{\it Second-Generation HTS Conductors}, edited by A. Goyal, 
(Kluwer Academic Publishing, Boston, 2005). 
\bibitem{Foltyn}
S.R. Foltyn $et.$ $al.$, Nat. Mater. {\bf 6}, 631 (2007), 
and refers. herein.
\bibitem{Kang} 
S. Kang $et.$ $al.$, Science {\bf 311}, 1911 (2006).
\bibitem{Claudia}
C. Cantoni $et.$ $al.$, to appear on ACS Nano (2011).
\bibitem{Baca2}
F.J. Baca $et.$ $al.$, Appl. Phys. Lett. {\bf 94}, 102512 (2009)
\bibitem{Mele}
P. Mele P $et.$ $al.$, Supercond. Sci. Technol. {\bf 21}, 015019 (2008).
\bibitem{Maiorov}
B. Maiorov $et.$ $al.$, 2009 Nature Mater. {\bf 8}, 398 (2009).
\bibitem{Rodriguez}
J.P. Rodriguez, P.N. Barnes, and C.V. Varanasi, Phys. Rew. B{\bf 78}, 
052505 (2008). 
\bibitem{Khachaturyan}
A.G. Khachaturyan, {\it Theory of Structural Transformations in Solids},
(John Wiley \& Sons, New York, 1983).
\bibitem{Bimberg} 
D. Bimberg, M. Grundmann, and N.N. Ledentsov, 
{\it Quantum Dot Heterostructures}, (John Wiley \& Sons, New York, 1999).
\bibitem{Roitburd}
A.L. Roitburd, in {\it Solid State Physics, Advances in Research and 
Applications}, edited by H. Ehrenreich, F. Seitz, and D. Turnbull 
(Academic, New York, 1978), Vol. 33, p. 317.
\bibitem{Bruinsma}
R. Bruinsma and A. Zangwill, J. Physique {\bf 47}, 2055 (1986).
\bibitem{Ipatova}
I.P. Ipatova, V.G. Malyshkin, and V.A. Shchukin, 
J. Appl. Phys. {\bf 74}, 7198 (1993).
\bibitem{Ledentsov}
N.N. Ledentsov $et.$ $al.$, Phys. Rew. B{\bf 54}, 8743 (1996).
\bibitem{Artemev}
J. Artemev, Y. Jin, and A.G. Khachaturyan, 
Acta  Mater. {\bf 49}, 1165 (2001).
\bibitem{Slutsker1}
J. Slutsker, I. Levin, J. Li, A. Artemev, and A.L. Roytburd,
Phys. Rev. B{\bf 73}, 184127 (2006).
\bibitem{Slutsker2}
J. Slutsker, A. Artemev, and A.L. Roytburd,
Phys. Rev. Lett. {\bf 100}, 087602 (2008).
\bibitem{Goyal}
A. Goyal $et.$ $al.$, Supercond. Sci. Technol. {\bf 18}, 1533 (2005). 
\bibitem{Landau}
L.D. Landau and E.M. Lifshitz, {\it Theory of Elasticity}, 3rd ed. (1986).
\bibitem{Kuzel}
P. Kuzel $et.$ $al.$, J. Phys.: Condens. Matter {\bf 13}, 167 (2001).
\bibitem{Dieguez}
O. Di\'{e}guez, K.M. Rabe, and D. Vanderbilt, Phys. Rev. B{\bf 72}, 
144101 (2005).
\bibitem{Bouhemadou}
A. Bouhemadou and K. Haddadi, Solid State Sciences {\bf 12}, 630 (2010).
\bibitem{Poindexter}
E. Poindexter and A.A. Gardini, Phys. Rev. {\bf 110}, 1069 (1958).
\bibitem{XXX}
W. Si, H.C. Li and X.X. Xi, Appl. Phys. Let. {\bf 74}, 2839 (1999).
\bibitem{Varanasi}
C.V. Varanasi $et.$ $al.$, J. Appl. Phys. {\bf 102}, 063909 (2007).
\end{thebibliography}
\end{document}